\documentstyle[12pt,aps,prb]{revtex}

\newcommand{\p}{\Psi_{\uparrow}}
\newcommand{\q}{\Psi_{\downarrow}}
\newcommand{\r}{\Phi_{\uparrow}}
\newcommand{\m}{\Phi_{\downarrow}}
\newcommand{\be}{\begin{equation}}
\newcommand{\ee}{\end{equation}}  
\newcommand{\ep}{\varepsilon}
\begin{document}

\draft

\title{ Magnetic Impurity in the two-dimensional    Heisenberg Antiferromagnet}

\author{V.N. Kotov, J. Oitmaa, and O. Sushkov}
\address{ School of Physics, University of New South Wales, Sydney 2052,
Australia}

\maketitle
\begin{abstract}

The two-dimensional    Heisenberg  model at zero temperature
with a quantum S=1/2 impurity spin, coupled to one site, is studied.
 The ground state properties of the model are calculated 
 for large coupling, in order to study the impurity - host spin local singlet
formation  
  and  related local suppression of the magnetization  in the host lattice.
Analytic results are obtained by developing  
perturbation theory  around the exactly solvable two-body (impurity plus
neighbor)
problem and compared with results obtained by exact diagonalization
of small clusters. 
We find that perfect screening of the impurity is only
achieved for infinitely large interaction, while at intermediate
coupling the local magnetization is  suppressed, but non-zero.

\end{abstract}

\narrowtext

\newpage
\section{Introduction}

The problem of  magnetic impurities  interacting with a system
of strongly correlated  electrons  has attracted a lot  of
interest recently, mainly due to the experimental discoveries
of  the high-$T_{c}$ superconductors and new heavy fermion 
compounds. In the field of the high-$T_{c}$  materials,
the parent compounds are known to be
two-dimensional (2D) antiferromagnetic (AFM)
Mott-Hubbard insulators which are driven to  a superconducting
state by doping (e.g. with holes)\cite{Manousakis,Dagotto}. 
Even though the holes can  hop, thus destroying the
AFM long-range order (LRO) and causing the development of
superconducting pairing, the extreme limit of static holes
is also believed to have physical relevance.
More generally, the effect of  local perturbations on the AFM
order is an interesting problem by itself. 
Several problems in 2D have been studied - static vacancy (missing site)
\cite{Bulut}, an impurity spin with an on-site \cite{Nagaosa}
and sublattice symmetric \cite{Jaan,us} coupling, as well
as an isolated ferromagnetic bond \cite{Lee,us}. Impurity spins,
coupled to 1D antiferromagnets have also been considered, by applying
bosonization techniques \cite{Onedim}.
 
 From the perspective of   the heavy fermion physics, in light
of some recent experimental observations, it is important to
take into account strong correlations between the conduction
electrons, interacting with an on-site magnetic impurity.
As a starting point, a model at half filling, with 
 a large Hubbard on-site repulsion, leading to
 the freezing of
the conduction degrees of freedom, has been proposed \cite{Fulde2}.
The problem  reduces to an impurity spin, coupled to a
 Heisenberg antiferromagnet. In 2D, for $T=0$  the problem has been
studied by using the linear spin-wave approximation 
(LSWA) \cite{Fulde2}. 
For $T \neq 0$, when LRO is absent in 2D,  the Schwinger boson mean-field theory
was applied  
 \cite{Murayama}.
Also, the one-dimensional version of the problem has been
studied numerically as a toy model \cite{Fulde1,Fulde3}.

In the present work we consider a magnetic impurity, coupled
via an on-site Kondo coupling to a 2D quantum Heisenberg
antiferromagnet at $T=0$. Our main  goal is the investigation
of the interplay between LRO and local screening of the
impurity spin (Kondo effect). Since LRO is present in 2D,
with a staggered moment reduced to about 61 \% from its classical
value, the Kondo coupling, necessary to induce considerable
impurity screening is rather large. Previous works have used  
the LSWA with subsequent perturbation theory in the Kondo interaction
\cite{Fulde2,Murayama} - an approach not suitable for the study of strong 
interactions.
 To get insight into the strong-coupling regime,
we observe that due to the local character of the perturbation
 it is sufficient first to diagonalize exactly  the two-body system,
consisting of the
impurity and directly coupled substrate spin,
and then treat the remaining interaction with the AFM environment
in perturbation theory. Technically  we  work with a  two-particle 
Green's function which we use to construct the ground state of the system.  
 The rest of the paper is
organized as follows. Our approach is described in Section II.
We also compare our analytical results with results obtained from
 exact diagonalization studies. The numerical procedure is outlined
in Section III. Section IV contains summary of our results and discussion.

\section{Impurity spin in the Heisenberg antiferromagnet.}

Consider the two-dimensional spin-1/2 Heisenberg model
 at zero temperature with a single, spin-1/2 magnetic impurity:

\begin{equation}
  H = J\sum_{<i,j>}\vec{S}_{i} . \vec{S}_{j} + 
K \vec{S}_{0}.  \vec{\sigma}
\end{equation}
All the couplings
are antiferromagnetic $J > 0, K > 0$. 
We set $J=1$ from now on. In the   first term in (1) the
summation is over nearest neighbors on a square lattice. The second
term represents an S=1/2  spin $\vec{\sigma}$ coupled to the site
$i=0$ of the host lattice  via  the Kondo 
coupling $K$.
 It is known \cite{Manousakis} that for $K=0$  there is
LRO in the ground state of $H$ and the  staggered magnetic
moment $m \equiv <S_{i}^{z}> = 0.303$. The Kondo interaction favors 
singlet formation and thus suppresses the magnetization locally, leading
also to screening of the impurity spin.  

In order to treat the strong coupling case $K>1$ we start by
rewriting   the Hamiltonian
as
\begin{equation}
H = H_{sw} + \left\{K \vec{S}_{0}.
\vec{\sigma} + 
 S_{0}^{z}\sum_{i=1}^{4}S_{i}^
{z} \right\}
+ \left\{ \frac{1}{2} S_{0}^{+}
\sum_{i=1}
^{4}S_{i}^{-} +\mbox{ h.c.}
\right\}, 
\end{equation}

\begin{equation}
 H_{sw} = \sum_{\bf k} \varepsilon_{\bf k}
(\alpha_{\bf k}^{\dagger}
\alpha_{\bf k} + \beta_{\bf k}^{\dagger} \beta_{\bf k} ).
\end{equation}
The spins $\vec{S}_{i}, i=1-4$ are the nearest neighbors of
$\vec{S}_{0}$ (see Fig.1.) and are assumed to belong to sublattice B (spin down).
Here  $\beta_{\bf k}$ and $\alpha_{\bf k}$ are the usual spin-wave
operators with dispersion \cite{Manousakis}: 

\begin{equation}
\varepsilon_{\bf k} = 2\sqrt{1- \gamma_{\bf k}^{2}}, \\\
 \gamma_{\bf k} = \frac{1}{2}(\mbox{cos}(k_{x}) +
 \mbox{cos}(k_{y})).
\end{equation}
In (2) we have explicitly separated the host spin-impurity and
host spin-AFM background interaction.
We  have assumed that the exclusion of one spin
(the one at $i=0$) does not influence the spin-wave Hamiltonian.
 This is valid in the one-loop approximation (see below)
which we use in the present work. However, if  one wants to
go beyond the lowest order (i.e. to higher loops),   
the influence of the $i=0$ 
perturbation on the spin-wave spectrum has to be taken into
account.

Since the impurity spin disturbs the AFM background only
locally, we can make the mean-field substitution
$S_{0}^{z}S_{i}^{z} \rightarrow S_{0}^{z}<S_{i}^{z}> =  S_{0}^{z}m $.
The  terms in the first curly brackets in (2) then can be diagonalized exactly,
while the  ones in the second curly brackets 
will be treated as a
perturbation. 
 For the spins $\vec{S}_{0}$ and $\vec{\sigma}$ we introduce
the fermionic representation:

\be
S^{+}_{0} = \p^{\dagger} \q, \ \  
S^{z}_{0} = \frac{1}{2}(\p^{\dagger}\p - \q^{\dagger} \q),
\ee
where $\p^{\dagger}$ and $\q^{\dagger}$ create
(acting on their common vacuum) a $S_{0}^{z}$ component
$1/2$ and $-1/2$ respectively.
In order for the two fermions to represent  a spin-1/2 operator,
they have to satisfy the constraint: \mbox{ $\p^{\dagger}\p + 
\q^{\dagger}\q =1$}, i.e. the physical states have only one fermion.
The operators representing $\vec{\sigma}$ via a formula
identical to (5) are denoted by $\r,\m$.
Using this representation and the usual spin wave expansion
for the spins $\vec{S}_{i}, i=1-4$,  the Hamiltonian becomes:

\be
H = H_{sw} + H_{0} + H_{int},
\ee

\be
H_{0} = 2m (\q^{\dagger}\q - \p^{\dagger} \p) +
\frac{K}{4}(\p^{\dagger}\p - \q^{\dagger}\q)
(\r^{\dagger}\r - \m^{\dagger}\m) +
\frac{K}{2}(\p^{\dagger}\q \m^{\dagger}\r +
\mbox{h.c.}),
\ee

\begin{equation}
H_{int} 
 =
 \sqrt{\frac{8}{N}} \p^{\dagger}\q \sum_{\bf k}\gamma_{\bf k}
(u_{\bf k} \beta_{\bf k} + v_{\bf k}\alpha_{-\bf k}^{\dagger})
+ \mbox{ h.c.},
\end{equation}
where $N$ is the total number of lattice sites, and 
 $u_{\bf k} = \sqrt{\frac{1}{2} + \frac{1}{\varepsilon_{\bf k}}} $,
  $v_{\bf k} = -\mbox{sign}(\gamma_{\bf k})
\sqrt{-\frac{1}{2} + \frac{1}{\varepsilon_{\bf k}}} $
are the parameters of the Bogoliubov transformation. The ${\bf k}$
sums in (8) as well as all future formulas 
 are over half of the first Brillouin zone, $0<k_{x},k_{y}<\pi$.

We have previously applied an approach, similar to the one 
that leads to the effective Hamiltonian (6), to 
study locally frustrating defects  in quantum
antiferromagnets \cite{us}. The following treatment is
also closely related to the one used by us in Ref.[6].

 In order to diagonalize $H_{0}$
it is convenient to define the two-particle matrix Green's
function 
$\hat{G}_{\mu \nu}$:

\be
\hat{G}_{\mu \nu}(t) = -i \left(
\begin{array}{cc}
 <T(\p(t) \m(t) \p^{\dag}(0)\m^{\dag}(0))>& <T(\p(t) \m(t)\q^{\dag}(0)
\r^{\dag}(0))>\\
<T(\q(t) \r(t)\p^{\dag}(0)\m^{\dag}(0))>&  <T(\q(t) \r(t) \q^{\dag}(0)
\r^{\dag}(0))> 
\end{array}
\right).
\ee
As  (9) suggests, 
the diagonal elements (11 and 22) correspond to the
two-particle 
states $|1> \equiv  |\uparrow\downarrow>$ and
 $|2> \equiv  |\downarrow \uparrow>$, respectively, where
the first arrow represents the spin  $\vec{S}_{0}$ and
the second one - the spin  $\vec{\sigma}$, or, equivalently
$ |\uparrow\downarrow>= \p^{\dagger}\m^{\dagger}|0>,
|\downarrow \uparrow>=\q^{\dagger}\r^{\dagger}|0>$, where
$|0>$ is the fermionic vacuum.
The off-diagonal components represent  transitions between
these two states. For future purposes we define
also the states: $|3> \equiv  |\uparrow \uparrow>$ and
$|4> \equiv  |\downarrow\downarrow>$.
 The unperturbed  Green's
functions (corresponding to $H_{0}$) are:

\be
G_{11,22}(\omega) = \frac{1}{\omega + K/4 \pm 2m +i\delta}, \ \
G_{12}(\omega) = G_{21}(\omega)= \frac{1}{\omega - K/2 + i\delta}.
\ee
Next, we evaluate the self-energy corrections to (10)
 to lowest, one-loop 
 order  in perturbation
theory  with respect to  $H_{int}$. 
Evaluating the diagram of Fig.2a, we obtain:

\begin{equation}
\Sigma_{11}(\varepsilon)= i 
\frac{8}{N}\sum_{\bf k}\gamma^{2}_{\bf k}
u^{2}_{\bf k} \int D({\bf k},\ep')G_{4}(\ep - \ep')
\frac{d\ep'}{2\pi} =  
\frac{8}{N}\sum_{\bf k} \frac{\gamma_{\bf k}^{2}
 u_{\bf k}^{2}}{
\varepsilon -K/4 -2m - \varepsilon_{\bf k}}.
\end{equation}
Here $ D^{-1}({\bf k},\omega) = \omega
- \ep_{\bf k}$ is the spin-wave Green's function,
and $G_{4}^{-1}(\omega) =
 \omega - K/4 - 2m$ corresponds to the two-particle
state $|4>$.
Analogous calculation gives (see Fig.2b):

\begin{equation}
\Sigma_{22}(\varepsilon)=
\frac{8}{N}\sum_{\bf k} \frac{\gamma_{\bf k}^{2}
 v_{\bf k}^{2}}{
\varepsilon -K/4 +2m - \varepsilon_{\bf k}},
\end{equation}
and $\Sigma_{12}=\Sigma_{21} =0$.
Higher loop corrections to the self-energy  can also be 
taken into account. Let us mention that
vertex corrections do not exist due to spin conservation
(reflected by the structure of the interaction (8)). The only
remaining diagrams are the "rainbow" ones \cite{remark}.
 We find that their contribution
is  small compared to the dominant one of Eq.(11) and Eq.(12).
Thus we restrict ourselves to the one-loop  order.

Since the $z$ component of the total spin is conserved, 
the  wave function of the spins
$S_{0}$ and $\sigma$ is spanned by the states
$|1>$ and $|2>$, corresponding to $S_{0}^{z} + \sigma^{z} =0$. 
The equation for the effective energy level
$\ep^{*}$ is:

\be
\left| 
\begin{array}{cc}
 \ep^{*}+K/4+2m - \Sigma_{11}(\ep^{*})& K/2\\
 K/2&\ep^{*}+K/4-2m - \Sigma_{22}(\ep^{*})
\end{array} 
\right| =0.
\ee
The correct normalized eigenstate is of the form
$|12> = \mu_{1}|1> + \mu_{2}|2>$, where $(\mu_{1},\mu_{2})$ is an eigenvector
of the matrix in (13), and $\mu_{1}^{2}+\mu_{2}^{2}=1$.
The ground state wave function of the two-particle - spin-wave system 
  can be written as

\be
|G> = \sqrt{Z}|12>|\mbox{sw}> + 
 \sum_{\bf k}B_{\bf k} 
|\downarrow \downarrow> \beta_{\bf k}
^{\dagger}|
\mbox{sw}> +  
 \sum_{\bf k} A_{\bf k}
|\uparrow \uparrow> \alpha_{\bf k}
^{\dagger}|
\mbox{sw}>.
\ee
We have defined:
\be
B_{\bf k} = \sqrt{\frac{8}{N}} \frac{\mu_{1}\gamma_{\bf k}u_{\bf k}}
{\ep^{*} - K/4 -2m- \ep_{\bf k}}, A_{\bf k} = \sqrt{\frac{8}{N}}
\frac{\mu_{2}\gamma_{\bf k}v_{\bf k}}{\ep^{*} -K/4 +2m - \ep_{\bf k}}, 
\ee
and $|\mbox{sw}>$ is the spin-wave vacuum, i.e.
$\alpha_{\bf k}|\mbox{sw}>=\beta_{\bf k}|\mbox{sw}>=0$.
The first term in (14) is  the coherent  part of the wave function
while  the rest is the contribution of the
intermediate states.
The normalization factor $Z$ is defined as:

\be
Z = 1 +\mu_{1}^{2} \left( \frac{\partial  \Sigma_{11} }{\partial\ep}
\right)_{\ep=\ep^{*}}
+ \mu_{2}^{2} \left( \frac{\partial  \Sigma_{22} }{\partial\ep}
\right)_{\ep=\ep^{*}}.
\ee
The average of any spin operator  in the state $|G>$ can be
computed by using the explicit form (14).
The results for the magnetization of the impurity spin and its 
neighbor as well as their spin-spin correlation function are:

\be
M(0) \equiv  <S_{0}^{z}> = \frac{\mu_{1}^{2} -\mu_{2}^{2}}{2}
Z + \frac{1}{2} \sum_{\bf k}\left( A_{\bf k}^{2} -  B_{\bf k}^{2}
\right),  
\ee

\be
M(\sigma) \equiv  <\sigma^{z}> =  \frac{\mu_{2}^{2} -\mu_{1}^{2}}{2}
Z + \frac{1}{2} \sum_{\bf k}\left( A_{\bf k}^{2} -  B_{\bf k}^{2}
\right),
\ee

\be
C(\sigma,0) \equiv <\vec{\sigma}.\vec{S}_{0}> 
 = \frac{1-2Z}{4} +  \mu_{1} \mu_{2} Z.
\ee

In order to find the spin-spin correlation between the host spin
$\vec{S}_{0}$ and its nearest neighbor $\vec{S_{1}}$ of the
host lattice (see Fig.1), we need
to find  how  $\vec{S_{1}}$ acts on the spin-wave states
$|\mbox{sw}>, \alpha_{\bf k}^{\dagger}|\mbox{sw}>,
\beta_{\bf
k}^{\dagger}|\mbox{sw}>$, which appear in $|G>$. This is accomplished
by using the  Holstein-Primakoff representation $S_{1}^{z} = -1/2
+ b_{1}^{\dagger} b_{1}, S_{1}^{+} =  b_{1}^{\dagger}$, and
performing the Bogoliubov transformation
$b_{\bf k} = u_{\bf k} \beta_{\bf k} + v_{\bf k} \alpha_{- \bf k}^{\dagger}$.
The final result is

\begin{equation}
C(1,0)  \equiv <\vec{S}_{1}.\vec{S}_{0}>=
<S_{0}^{z}> \left \{ -\frac{1}{2}  + 
\frac{2}{N} \sum_{\bf k} v_{\bf k}^{2} \right\} + 
\frac{1}{2}\left[
 \mu_{1}^{2}\sqrt{Z} \Sigma_{11} + \mu_{2}^{2}\sqrt{Z} \Sigma_{22}
\right].
\end{equation}
In  Eq.(20) the self-energies are evaluated at
$\ep = \ep^{*}$. The expression in the curly brackets is the LSWA magnetization 
 $m=0.303$.
 In order to compute (17-20) all the lattice sums
as well as the solution of Eq.(13) have to be calculated  numerically.
The results are summarized in Figures 3 and 4.
 
Notice that in the approximation we have adopted for
the Hamiltonian Eq.(6), the magnetization of the spins 1,2,3,4
(see Fig.1) is
$m=0.303$ (since they are part of the spin-wave background), and does not
depend on $K$.
In order to calculate, e.g. $<S_{1}^{z}>$ more accurately, 
 one needs to diagonalize exactly        
 the system of three spins $S_{0},
S_{1},\sigma$ and treat all the rest of the spins in the linear spin-wave
approximation. Conceptually the calculation is very similar
to the one presented above. However, since the technical details are more
involved and not particularly illuminating, 
we do not describe them  here \cite{remark1}. The results for  $<S_{1}^{z}>$
are given in Table.I. The diagonalization of the three-body system
is expected to produce  also more accurate results for $C(1,0)$. We have
found, however, that the difference between the three-body calculation
and Eq.(20) is numerically very small.        

In our technique the    magnetization of the
unperturbed Heisenberg antiferromagnet ($K=0$) differs from
the LSWA value $m=0.303$ and depends on the size of the "cluster" -
the number of spins that are exactly diagonalized. We find 
the magnetization to be 0.272 if the cluster consists of one spin
($M(0)$  from Fig.3.) and 0.318 for a cluster of two spins 
($<S_{1}^{z}>(GF)$ 
from Table.I). We generally expect that the accuracy of our method
will increase as the size of the cluster increases  and higher
orders of perturbation theory in the cluster-AFM background interaction
are taken into account.

The results for the magnetizations as  functions of $K$
 (Fig.3) are consistent with
the physically expected behavior. As $K$ increases, the impurity
spin becomes gradually screened, while the magnetization of the
nearest substrate spin decreases. Perfect singlet formation between
$\vec{S}_{0}$ and $\vec{\sigma}$ is achieved only in the limit
of infinitely large $K$.  The  behavior of the correlation functions
(Fig.4)
supports the above picture. The maximum value  of 0.75 for the correlation
function $C(\sigma,0)$ is gradually approached with the increase of $K$.
Notice that  $<S_{1}^{z}>$ (Table.I) at any $K\neq0$ is
larger than the value at $K=0$. This  means that the quantum fluctuations
have effectively decreased at that site.  
 Such behavior was also observed in the 2D Heisenberg model
with a missing site \cite{Bulut}.

For small value of the Kondo coupling $K<1$ our results for the magnetization
are consistent with previous calculations of Igarashi, Murayama, and Fulde
\cite{Fulde2} in this regime, who  used LSWA in combination with perturbation theory
in $K$. We emphasize, however, that our calculation was specifically
designed to treat the strong-coupling limit $K>1$.
 Similar  results  to ours have been reported in the one-dimensional version
of the model by Igarashi {\it et al.}\cite{Fulde1}. 
They studied numerically the behavior of the quantities, defined in
Eqs.(17-20). From  comparison of our curves with theirs, we
conclude that the tendency for local singlet formation is more
pronounced  for  a Heisenberg chain, which is expected, since LRO is not present
in 1D.

\section{Exact diagonalization studies.}

We  have  also studied the system  by exact
diagonalization of small clusters. Traditionally
periodic boundary conditions (PBC) are used \cite{Jaan1}.
 However we have found in the study of
 related problems, involving locally frustrating spin defects \cite{Jaan}
that the use of PBC may lead to spurious discontinuities in
the observables. We have attributed this to finite size effects \cite{us}.
Instead, in our previous work \cite{us}  we  applied
a staggered magnetic field in the $z$ direction to spins on
the boundary of the cluster (see Fig.1). The same procedure is used
here. We have chosen a cluster of $N=18+1$ spins and the boundary
field to give $<S^{z}>=0.3$ on the boundary spins. The advantage of such
an approach is that the sublattice symmetry is broken which allows us
to compute single spin averages and to distinguish between longitudinal
and transverse correlations. Our results are summarized in Table.I and
 Figures 3 and
4, for comparison with the analytic approach. Due to the asymmetry
of the cluster the correlation function $C(1,0)$ is
computed as $ [C(1,0) + C(2,0) + C(3,0) + C(4,0)]/4$ (see Fig.1). Analogous
symmetrization was used for $<S_{1}^{z}>$.
We find a very good qualitative agreement between the analytical and exact
diagonalization 
results.


Our numerical procedure  also breaks the translational invariance
of the cluster and will, in general, lead to increased finite size
effects and slow convergence to the bulk limit. 
To achieve complete self-consistency, the extrapolation to the 
thermodynamic limit has to be performed, by adjusting simultaneously  the boundary
field to be equal to the magnetization of the spins inside the cluster. 
This certainly would increase even further the numerical agreement
between the analytical and numerical results.

\section{Summary and discussion.}

To summarize, we have studied the competition between LRO and local singlet
formation in the 2D Heisenberg model at $T=0$, coupled to a magnetic impurity via
an on-site Kondo term. We were particularly interested in developing
a formalism to treat large Kondo couplings, since it is physically
clear, that due the presence of LRO in 2D, only large couplings
can lead to substantial impurity screening and local suppression of the
magnetization. We find that the local singlet is formed asymptotically.       
At intermediate couplings the magnetization sustains a non-zero value,
which gradually decreases as the coupling increases. This picture
is supported by the two methods we have used: 1.) An analytic approach,
which treats exactly the system, consisting of the
impurity  and the directly coupled to it host spin, and
takes  into account the interaction with the AFM background perturbatively,
and 2.) Exact diagonalization of small clusters.

A related problem is the behavior of a Kondo moment in a system
without LRO in the ground state (non-zero gap in the magnon
energy spectrum). Destruction of LRO can be achieved
e.g. by doping, or by introduction of additional spin interactions.
The lack of LRO would lead to different behavior, compared to the one
found in this paper, since the impurity could become completely
screened at intermediate couplings. An interesting question is
the existence of impurity induced bound states in the gap.
Such states have been found  in the $S=1$ Heisenberg chain \cite{Sorensen}, 
and in s-wave (as well as d-wave under certain conditions) 
superconductors with magnetic impurities \cite{Salkola}.
We plan to address these issues in future work.

\acknowledgments
We would like to acknowledge the financial support of
the Australian Research Council.

\begin{table}
\caption{ Magnetization  on site 1
calculated by the Greeen's function (GF)
technique  and
 by exact diagnalization (ED)
of N=18+1 cluster.}
\label{table1}
\begin{tabular}{c|cccccc}
K  & 0 & 1 &  2  & 3  & 4  & 5 \\ \hline
$|<S_{1}^{z}>|$(GF) & 0.318 &0.327 & 0.334 &0.330 & 0.329 &0.328 \\
$|<S_{1}^{z}>|$(ED)& 0.3717 &0.3804 & 0.3822 &0.3820 & 0.3814 &0.3809 \\
\end{tabular}
\end{table}

\vspace{1cm}

FIGURE CAPTIONS

\vspace{0.5cm}

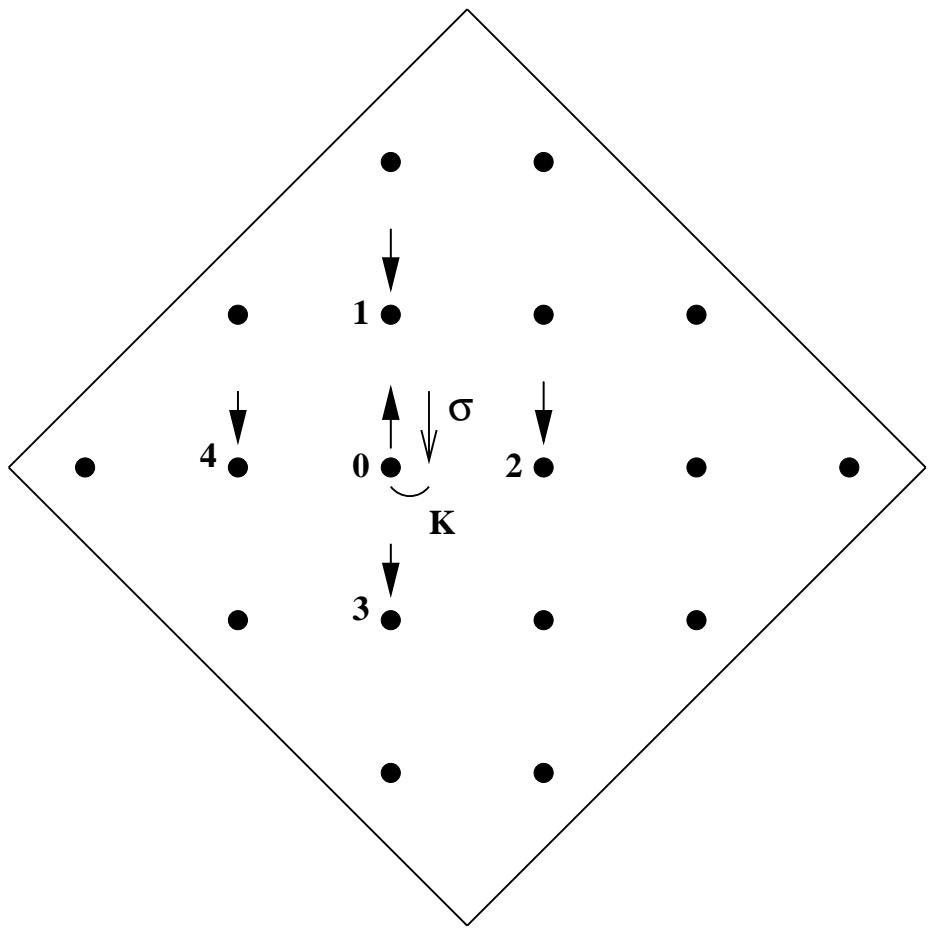
\begin{figure}
\caption{ Schematic representation of the system. The black dots
are the host spins, with interaction J=1 between them.
A cell, containing N= 18 + 1 sites, used for   exact diagonalization,
is shown. 
}
\label{fig1}
\end{figure}

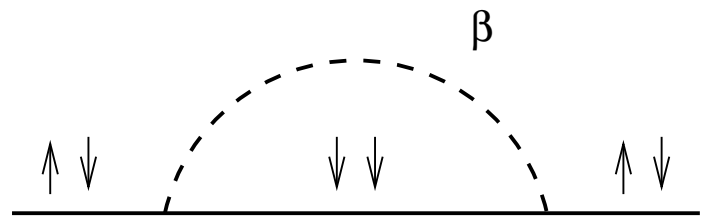
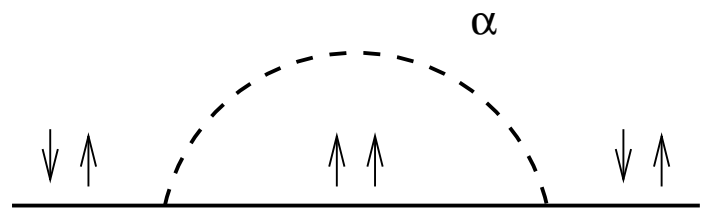
\begin{figure}
\caption{ One-loop diagrams, contributing to the two-particle
Green's function, Eq.(9). Solid and dashed lines represent
fermion and spin-wave Green's functions, respectively.
}
\label{fig2}
\end{figure}

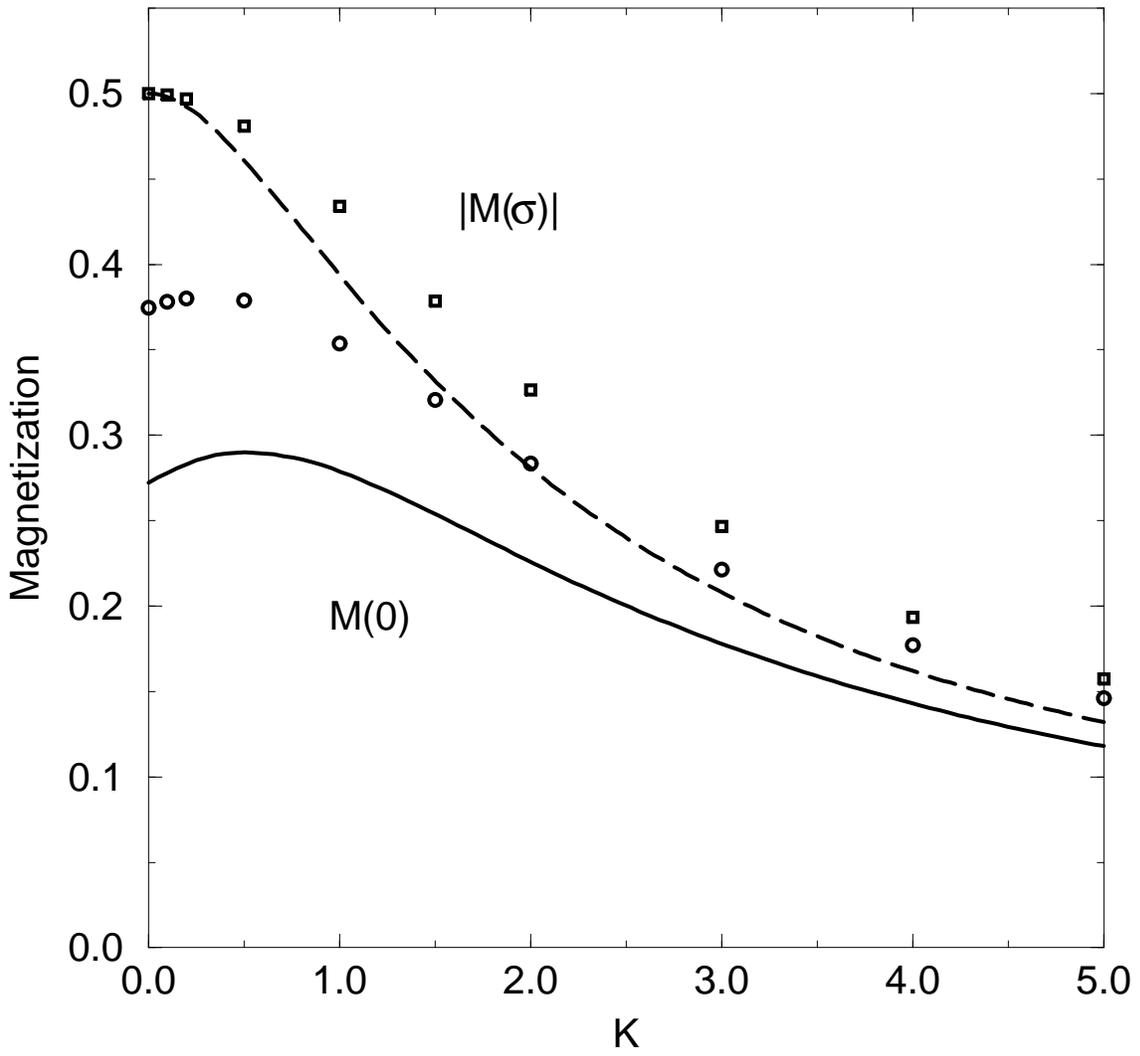
\begin{figure}
\caption{The impurity spin average $M(\sigma)$ and the magnetization
of the neighboring host spin $M(0)$. The solid and dashed curves are
calculated from Eq.(17) and Eq.(18), while the circles and squares are
the corresponding exact diagonalization results.
}
\label{fig3}
\end{figure}

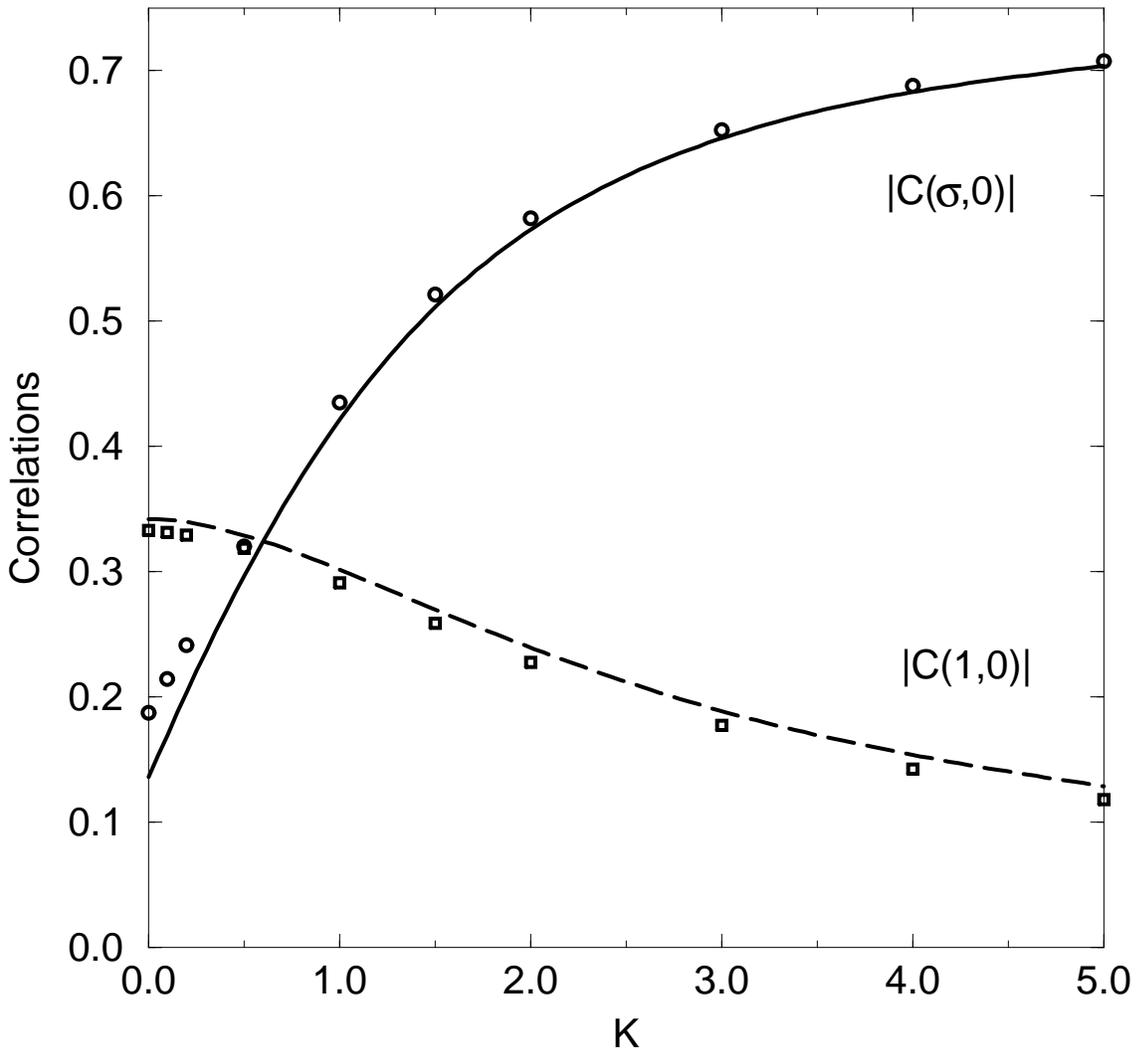
\begin{figure}
\caption{ Spin-spin correlation functions as defined by Eq.(19) and Eq.(20).
The  solid and dashed line are obtained by numerical evaluation of Eqs.(19-20).
The open symbols are the corresponding exact diagonalization results.
}
\label{fig4}
\end{figure}


\begin{references}
\bibitem{remark} The absence of vertex corrections in
this problem  is analogous to the same effect in the t-J model
\cite{Dagotto},
and follows from the form of the fermion - spin-wave vertex.
\bibitem{remark1} For details of a  similar calculation, involving
three spins, see Ref.[6], where the problem of an
external spin,
coupled symmetrically to the two sublattices, is studied.
\end{references}
\end{document}